\documentclass[12pt]{article}
\makeatletter
\def\fmslash{\@ifnextchar[{\fmsl@sh}{\fmsl@sh[0mu]}}
\def\fmsl@sh[#1]#2{%
  \mathchoice
    {\@fmsl@sh\displaystyle{#1}{#2}}%
    {\@fmsl@sh\textstyle{#1}{#2}}%
    {\@fmsl@sh\scriptstyle{#1}{#2}}%
    {\@fmsl@sh\scriptscriptstyle{#1}{#2}}}
\def\@fmsl@sh#1#2#3{\m@th\ooalign{$\hfil#1\mkern#2/\hfil$\crcr$#1#3$}}
\makeatother
\global\arraycolsep=2pt 
\input{epsf}
\begin{document}
\thispagestyle{empty}
\rightline{TTP01-34}
\rightline{hep-ph/0201136}
\rightline{December 2001}
\bigskip
\boldmath
\begin{center}
{\bf \Large
Reparametrization Invariance in}  \\[3mm]
{\bf \Large Inclusive Decays of Heavy Hadrons}
\end{center}
\unboldmath
\smallskip
\begin{center}
{\large{\sc  Francisco Campanario}} and  {\large{\sc Thomas Mannel}} 
\vspace*{2cm} \\
$^{(b)}$ {\sl Institut f\"{u}r Theoretische Teilchenphysik, \\
Universit\"{a}t Karlsruhe,  D--76128 Karlsruhe, Germany}
\end{center}
\begin{abstract}
\noindent
Reparametrization invariance is the invariance of the heavy mass limit
under small changes of the heavy-quark four velocity.  
We discuss the implications of this invariance for
non-local light cone operators, the matrix elements of which 
are relevant for the leading and subleading shape functions describing
differential rates for inclusive heavy-to-light transitions. 
\end{abstract}
\newpage
\section{Introduction}
Recent investigations of the heavy-mass expansion for heavy meson
decays made the role of certain non-local operators appearent. 
This type of operators appears in the context of
differential rates for heavy-to-light decays, where e.g. the photon
spectrum of $B \to X_s \gamma$ is expressed in terms of the so-called
shape function $f(\omega)$ \cite{shape} which is given as 
\begin{equation}
f(\omega) = \langle B(v) |\bar{h}_v \delta (\omega + (in\cdot D)) h_v
            | B(v) \rangle \, , 
\end{equation}
where $n$ is a certain light cone vector determined by the kinematics.
This expression, given 
as a matrix element of a non-local light cone operator,
corresponds to the leading term of a twist expansion of the
inclusive decay rates, in analogy to deep inelastic scattering.

The subleading terms have been investigated at tree level in \cite{BLM1}. 
While the leading term can be interpreted in terms of light-like Wilson lines
connecting two heavy quarks at ``light-cone time'' 0 and $t$, 
the subleading terms can be interpreted as light-like Wilson
lines with ``insertions'' of additional covariant derivatives, leading
to contributions suppressed by one power of $1/m_Q$. 

The analysis of the subleading terms beyond tree level has not yet
been performed. Including radiative corrections to the leading shape
function leads schematically to a rate of the form
\cite{KS}
\begin{equation}
d \Gamma = H \otimes J \otimes S
\end{equation}
where $\otimes$ denotes a convolution of the functions $S$, $J$ and $H$.
Here $H$ describes a hard, process dependent contribution, $J$ is a
``jetlike'' contribution, containing also the Sudakov logarithms \cite{BFS}, 
and $S$ describes the soft terms. For the subleading shape functions
a similar pattern of the radiative corrections is expected. 

In deriving the heavy mass expansion  from QCD
one introduces a velocity vector $v$ which is
the velocity of the hadron containing the heavy quark,
$v=p_{hadron}/M_{hadron}$. The heavy quark momentum  $p_Q$ inside
the heavy meson is decomposed 
into a large part $m_Q v$ and a residual part $k$, $p_Q = m_Q v + k$,
and the heavy mass expansion is constructed by expanding the amplitudes 
in the small quantity $k/m_Q$. From this point of view 
the velocity vector $v$ in HQET is an external variable, which
is not present in full QCD, and which is  only fixed up to terms of the
order $\Lambda_{QCD} / m_b$. Consequently, small reparametrizations of the
form $v \to v + \Delta$ with $\Delta = {\cal O} (\Lambda_{QCD} / m_b)$ 
should leave the physical results of the heavy mass expansion invariant. 

This so-called reparametrization invariance is known since the
early days of heavy quark effective theory (HQET) \cite{repara}
and its main feature is that it connects different
order of the $1/m_b$ expansion. Many applications of
reparametrization invariance have been studied, 
the most prominent of which is the non-renormalization of the kinetic
energy operator $\bar{h}_v (iD)^2 h_v$. 

The purpose of the present note is to exploit the
consequences of reparametrization invariance  for
the non-local operators appearing in the description of spectra
in heavy-to-light decays. The main result is that the number of unknown
functions appearing at order $1/m_b$ is reduced.

In the next section we discuss reparametrization invariance  in HQET, in
section~\ref{nonlocal} we discuss the light-cone operators and construct
reparametrization-invariant combinations of such operators. Finally we
consider applications and conclude. 

\section{Reparametrization Invariance}
We consider two versions of HQET  with two
different choices of the velocity vector $v$ and $v^\prime$ differing
by a small quantity $\Delta$\footnote{In the following we shall closely
follow the discussion given by Chen, second paper of \cite{repara}.}
\begin{equation} \label{velo}
v^2 = 1 \qquad v^{\prime 2}  = 1 = (v + \Delta)^2 = 1 + 2 v \cdot \Delta
+ {\cal O} (\Delta^2) \quad \mbox{ thus } \quad 
               v \cdot \Delta = 0   \, .
\end{equation}
If the change $\Delta$ in the velocity vector is of the
order  $\Lambda_{QCD}/m_Q$, the two versions of HQET have to be equivalent.

Constructing HQET from QCD involves a redefinition of the quark field
$Q$ of the form
\begin{equation}
Q = \exp (-im_Q v \cdot x) Q_v
\end{equation}
such that the covariant derivative acts as
\begin{equation}
i D_\mu Q =  \exp (-im_Q v \cdot x) (m_Q v + i D_\mu ) Q_v
\end{equation}
The left hand side corresponds to the full heavy quark
momentum which is not changed under reparametrization.
This implies for the change $\delta_R$ of the covariant derivative
acting on a the quark field $Q_v$ 
\begin{equation} \label{derivative}
\delta_R  (iD_\mu) = - m_Q \Delta_\mu  \, .
\end{equation}

In the following we have to develop a consistent scheme to count powers.
Defining the action to be ${\cal O}(1)$, we get that static heavy quark
field is ${\cal O}(\Lambda_{QCD}^{3/2})$. The covariant derivative as well
as the variation $\delta_R$ of the covariant derivative are
${\cal O}(\Lambda_{QCD})$, and the variaton of the heavy quark field
under reparametrization is 
\begin{equation} \label{fields}
\delta_R h_v = \frac{\fmslash{\Delta}}{2}
               \left[1 + \frac{i \fmslash{D}}{2 m_Q} \right] h_v
               + {\cal O}[\Lambda_{QCD}^{3/2} (\Lambda_{QCD}/m_Q)^3] \, .
\end{equation}
Note that the leading contribution originates from
the variation of the projector $P_+ = (\fmslash{v} + 1)/2$ and is of order 
$\Lambda_{QCD}^{5/2}/m_Q$

Equations (\ref{velo}), (\ref{derivative}) and (\ref{fields}) are the
reparametrization transformations of all relevant quantities needed to
exploit the consequences of this symmetry. 

Reparametrization invariance connects terms of different orders in the
$1/m_Q$ expansion. As an example we consider the HQET Lagrangian
\begin{eqnarray}
{\cal L} &=& {\cal L}_0 + {\cal L}_1 + \cdots 
         =  \bar{h}_v (iv\cdot D) h_v \\ \nonumber 
         && \qquad + \frac{1}{2 m_Q} \bar{h}_v (iD)^2 h_v
         - \frac{i}{2 m_Q} \bar{h}_v (i D_\mu) (i D_\nu) \sigma^{\mu \nu} h_v
         + {\cal O}(\Lambda_{QCD}^6 / m^2)  
\end{eqnarray}
with $h_v = P_+ h_v$ where $P_+ =(1+\fmslash{v})/2$.

The leading order term ${\cal L}_0$ is of order $\Lambda_{QCD}^4$, while
its variation is of order $\Lambda_{QCD}^5/m_Q$ 
\begin{equation}
\delta_R {\cal L}_0 =
\bar{h}_v (i \Delta \cdot D) h_v +{\cal O} [\Lambda_{QCD}^6/m_Q^2]
\end{equation}
Note that the leading term of the variation of the fields (\ref{fields})  
does not contribute since
\begin{equation} \label{lofield}
P_+ \fmslash{\Delta} P_+ = P_+ (v \cdot \Delta) = 0  
\end{equation}

The variation of the leading-order term
is compensated by the kinetic energy term, since
\begin{equation} \label{reparaL}
\delta_R \left( \bar{h}_v (iv\cdot D) h_v
         + \frac{1}{2 m_Q} \bar{h}_v (iD)^2 h_v \right) =
{\cal O}[\Lambda_{QCD}^6/m_Q^2]
\end{equation}
Relation (\ref{reparaL}) is preserved under renormalization which ensures
that the kinetic energy piece is not renormalized \cite{repara}

In a similar way one can obtain relations between higher order terms in the
Lagrangian and also for matrix elements. Again these relations do not change
under renormalization from which relations between renormalization constants
can be derived. 

\section{Light-Cone Operators} \label{nonlocal}
In inclusive decays the typical situation in which non-local light cone
operators are necessary is when a heavy quark decays into light 
particles and the energy spectrum of one of the outgoing particles
or a region of small invariant mass of a set of outgoing particle is
considered. The relevant kinematics for the case of an energy spectrum
for one outgoing particle are
\begin{equation}
p_b = m_Q v + k = q + p'  \qquad q^2 = 0 
\end{equation}
where $q$ is the momentum of the light particle for which the
energy spectrum is computed and $p'$ is the momentum of the rest of the
decay products. For the case of $B \to X_s \gamma$ one has at tree level
only one light-quark in the final state, leading to
$\delta (p^{\prime 2}) = \delta [( m_Q v + k - q)^2] $ as the spectral
function for the final state. For $B \to X_u \ell \bar{\nu}_\ell$ we have
at tree level a neutrino and a light quark in the final state, the spectral
function of which is proportional to
$\Theta (p^{\prime 2}) = \Theta [( m_Q v + k - q)^2]$. Thus the
lepton momentum plays the same role in  $B \to X_u \ell \bar{\nu}_\ell$
as the photon momentum in $B \to X_s \gamma$. Generically, the shape
function becomes relevant as soon as the spectral function of the remaining
particles is a step function close to $p^{\prime 2} = 0$. 

In order to describe the endpoint region of such an energy spectrum, i.e.
the region close to the maximal value of the energy $q \cdot v$, it  
is convenient to introduce light-cone vectors $n$ and $\bar{n}$ with
$n^2 = \bar{n}^2 = 0$ and $n \cdot \bar{n} = 2$,  
one of which is collinear with the momentum $q$ 
\begin{equation}
v_\mu  = \frac{1}{2} (n_\mu  + \bar{n}_\mu) \qquad
q_\mu  = \frac{1}{2} (n \cdot q) \bar{n}_\mu   \, .
\end{equation}
Using these relations we can write
\begin{equation}
m_Q v - q = \frac{m_Q}{2} n
          + \frac{1}{2} (m_Q - n \cdot q) \bar{n}  
\end{equation}
The endpoint region is now characterized by 
\begin{equation}
(m_Q - n \cdot q) \sim {\cal O} (\Lambda_{QCD}) 
\end{equation}
and a systematic expansion in $1/m_Q$ is performed.

The expansion close to the endpoint becomes an expansion in twist 
and cannot be performed in terms of local operators any more; rather
non-local light cone operators are needed. The
leading term has been  known for some time and the relevant
operators are \cite{shape}
\begin{eqnarray} \label{leading}
O_0 (\omega) &=& \bar{h}_v \delta (\omega + (in\cdot D)) h_v \\ 
P_0^\alpha (\omega) &=&  \bar{h}_v
\delta (\omega + (in\cdot D)) \gamma^\alpha \gamma_5 h_v
\end{eqnarray}
Note that $P_+ = (1+\fmslash{v})/2$ and $s_\mu = P_+ \gamma_\mu \gamma_5 P_+$
form a basis in the space of (two-component) spinors projected out by $P_+$.  

At subleading order the necesary set of operators can be
chosen as \cite{BLM1}
\begin{eqnarray} \label{subleading}
O_1^\mu (\omega) &=&
\bar h_v \left\{(i D^\mu),\delta(\omega + (i n \cdot D))\right\} h_v
\\ \nonumber
O_2^\mu(\omega) &=& i \, 
\bar h_v \left[(i D^\mu),\delta(\omega + (i n \cdot D))\right] h_v
\\ \nonumber
O_3^{\mu\nu}(\omega_1,\omega_2) &=&
\bar h_v \delta(\omega_2 + (in \cdot D))
     \left\{iD^\mu_\perp\, , \, iD^\nu_\perp \right\}
         \delta(\omega_1+ (i n \cdot D)) h_v
\\ \nonumber
O_4^{\mu\nu}(\omega_1,\omega_2) &=& i \, 
\bar h_v \delta(\omega_2 + (in \cdot D))
     \left[ iD^\mu_\perp\, , \, iD^\nu_\perp \right]
         \delta(\omega_1+ (i n \cdot D)) h_v
\end{eqnarray}
for the ``spin-independent'' operators and
\begin{eqnarray} \label{subleading-spin}
P_1^{\mu \alpha} (\omega) &=&
\bar h_v \left\{(i D^\mu),\delta(\omega + (i n \cdot D))\right\}
         \gamma^\alpha \gamma_5 h_v
\\ \nonumber
P_2^{\mu \alpha} (\omega) &=& i \, 
\bar h_v \left[(i D^\mu),\delta(\omega + (i n \cdot D))\right]
         \gamma^\alpha \gamma_5 h_v
\\ \nonumber
P_3^{\mu\nu \alpha}(\omega_1,\omega_2) &=&
\bar h_v \delta(\omega_2 + (in \cdot D))
     \left\{iD^\mu_\perp\, , \, iD^\nu_\perp \right\}
         \delta(\omega_1+ (i n \cdot D)) \gamma^\alpha \gamma_5 h_v
\\ \nonumber
P_4^{\mu\nu\alpha}(\omega_1,\omega_2) &=& i \, 
\bar h_v \delta(\omega_2 + (in \cdot D))
     \left[ iD^\mu_\perp\, , \, iD^\nu_\perp \right]
         \delta(\omega_1+ (i n \cdot D)) \gamma^\alpha \gamma_5 h_v
\end{eqnarray}
for the ``spin-dependent'' ones.

The (differential) rates are expressed in terms of convolutions of 
$\omega$-dependent Wilson coefficients with forward matrix elements of
these operators \cite{BLM1}
\begin{eqnarray} \label{rate}
d\Gamma &=& \int d \omega \left( C_0 (\omega) <O_0 (\omega)> +
            C_{0, \alpha}^{(5)} (\omega) <P_0^\alpha (\omega)> \right) \\
\nonumber
&& + \frac{1}{m_Q} \sum_{i=1,2}
       \int d \omega \left( C_{i, \mu}  (\omega) <O_i^\mu (\omega)> +
       C_{i, \mu \alpha}^{(5)} (\omega) <P_i^{\mu \alpha} (\omega)> \right)
\\ \nonumber
&& + \frac{1}{m_Q} \sum_{i=3,4}
       \int d \omega_1 \, d \omega_2
       \left( C_{i, \mu \nu }  (\omega_1, \omega_2)
             <O_i^{\mu \nu} (\omega_1, \omega_2)>  \right. \\ 
           && \qquad \qquad \qquad \qquad + 
       \left.     C_{i, \mu \nu \alpha}^{(5)} (\omega_1, \omega_2)
             <P_i^{\mu \nu \alpha} (\omega_1, \omega_2)> \right)
\nonumber \\ \nonumber && + \cdots
\end{eqnarray}
where $< .. >$ denotes the forward matrix element with $b$-Hadron states 
and the ellipses denote terms originating from time-ordered products with
higher order terms of the Lagrangian, which we do not need to consider here.  

In the following we want to discuss the implications of reparametriation
invariance for the non-local light-cone operators. Similar
to the case of local operators we shall derive reparametrization-invariant
combinations of operators containing different orders of the $1/m_Q$
expansion.
To investigate this we shall first compute the variation of the light
cone vectors under a reparametrization transformation, which means that
$v$ is varied according to (\ref{velo}) and $q$ is kept fixed. Expressing
the light-cone vectors in terms of $q$ and $v$ we get
\begin{equation}
n = \frac{1}{v \cdot q} [ 2(v \cdot q) v - q] \quad \mbox{and} \quad 
\bar{n} = \frac{1}{v \cdot q} q
\end{equation}
from which we can derive the variation $\delta_R$ under reparamatrization
\begin{eqnarray} \label{nnbar}
\delta_R n_\mu &=& \frac{\partial n_\mu}{\partial v_\alpha} \Delta_\alpha
= 2 \Delta_\mu + \bar{n}_\mu (\bar{n} \cdot \Delta) \\ \nonumber 
\delta_R \bar{n}_\mu &=&
\frac{\partial \bar{n}_\mu}{\partial v_\alpha} \Delta_\alpha
= - \bar{n}_\mu (\bar{n} \cdot \Delta)
\end{eqnarray}
Using this we can study the variation of
\begin{equation}
\hat{O}_0 (\omega) = \bar{h}_v \frac{1}{\omega + (in\cdot D)} \Gamma h_v
\end{equation}
which is of order $\Lambda_{QCD}^2$ since we have to count $\omega$ as
${\cal O}(\Lambda_{QCD})$. 
The imaginary part 
(by replacing $\omega \to \omega+i\epsilon$) of this expression
is either $O_0 (\omega)$ (for $\Gamma = P_+$) or $P_0^\alpha (\omega)$
(for $\Gamma = s^\alpha = P_+ \gamma^\alpha \gamma_5 P_+$).
From (\ref{derivative}) and (\ref{nnbar}) we get 
\begin{equation}
\delta_R (i n \cdot D) = -m_Q (n \cdot \Delta)
                       + (\bar{n} \cdot \Delta) (i \bar{n} \cdot D)
                       + 2 (i \Delta \cdot D)  
\end{equation}
and thus 
\begin{eqnarray} \label{deltaO0}
&& \delta_R \hat{O}_0 (\omega) =
   \bar{h}_v \{ \fmslash{\Delta} \, , \, \Gamma\}
              \frac{1}{\omega + (in\cdot D)} h_v \\  \nonumber  
&&  +   \bar{h}_v \frac{1}{\omega + (in\cdot D)} 
        [ m_Q (n \cdot \Delta)
                       - (\bar{n} \cdot \Delta) (i \bar{n} \cdot D)
                       - 2 (i \Delta \cdot D)] 
             \frac{1}{\omega + (in\cdot D)} \Gamma h_v \\ \nonumber && +
             {\cal O}[\Lambda_{QCD}/m_Q^2] 
\end{eqnarray}
where we have omitted terms of subleading order in $1/m_Q$
coming e.g.\ from the
variation of the heavy quark fields.

The first term vanishes due to (\ref{lofield})
and fact that $\Gamma$ is either $P_+$
or $s_\mu = P_+ \gamma_\mu \gamma_5 P_+$.
The second term contains a piece of order
$\Lambda_{QCD}^2$ (which is of the same order as $O_0 (\omega)$
itself) coming from the
variation of the covariant derivative, while all other terms in 
(\ref{deltaO0}) are of higher order.

We shall first discuss the variation of order $\Lambda_{QCD}^2$. 
This can be written as
\begin{eqnarray}
\delta_R \hat{O}_0 (\omega) &=& 
\bar{h}_v \frac{1}{\omega + (in\cdot D)}
              m_Q (n \cdot \Delta)
          \frac{1}{\omega + (in\cdot D)} \Gamma h_v
          + {\cal O} (\Lambda_{QCD}^3/m_Q) \\
&=& - \left(\frac{\partial}{\partial \omega} \hat{O}_0 (\omega) \right)
m_Q  (n \cdot \Delta) + {\cal O} (\Lambda_{QCD}^3/m_Q) 
\end{eqnarray}
which means that the ${\cal O} (\Lambda_{QCD}^2)$-variation can be
absorbed into a shift $\omega \to \omega - m_Q  (n \cdot \Delta)$.

In the following we assume that $\Delta$ does not have a light cone
component, i.e. we only consider $\Delta^\perp$ for which we have
$(n \cdot \Delta^\perp) = 0$. Note that this also implies
$(\bar{n} \cdot \Delta^\perp) =0$ due to (\ref{velo}). In this way
(\ref{deltaO0}) simplifies to
\begin{equation} \label{dperpO0}
\delta_R^\perp \hat{O}_0 (\omega) = 
             \bar{h}_v \frac{-2}{\omega + (in\cdot D)} 
             (i \Delta^\perp \cdot D) 
             \frac{1}{\omega + (in\cdot D)}  \Gamma h_v
             + {\cal O}(\Lambda_{QCD^4}/m_Q^2) \, .
\end{equation}

Our aim is to construct a reparametrization invariant, which is
equal to $O_0 (\omega)$ to leading order. The variation of $O_0 (\omega)$
of order unity as given in (\ref{dperpO0}), and a subleading
contribution is needed to compensate this variation. To construct this
invaraint, we first note that 
\begin{equation}
\delta_R^\perp \left((i n \cdot D) + \frac{1}{m_Q} (iD^\perp)^2 \right) = 0
\end{equation}
which means that
\begin{equation}
\left(\frac{1}
         {\omega + (i n \cdot D) + \frac{1}{m_Q} (iD^\perp)^2} \right)
\end{equation}
is an exact reparametrization invariant.

Furthermore, we may include higher order terms to construct a
reparametrization invariant field
\begin{equation}
H_v = h_v + \frac{(i\fmslash{D})}{2m_Q} h_v
                + \frac{1}{4 m_Q^2} (iD)^2 h_v + \cdots 
\, , \quad \delta_R^\perp H_v = 0
\end{equation}
which can be used to construct the reparametrization-invariant quantity 
\begin{equation}
\hat{R}_0 (\omega) = \bar{H}_v \left(\frac{1}
         {\omega + (i n \cdot D) +
                   \frac{1}{m_Q} (iD^\perp)^2} \right) \Gamma H_v
\end{equation}
where $\Gamma$ is again either $P_+$ or $s_\mu = P_+ \gamma_\mu \gamma_5 P_+$.

This formal expression can now be expanded to obtain the reparametrization
invariant combinaton of operators appearing in the twist expansion of
inclusive rates. Truncating the expansion yields operators for which 
where reparametrization invariance holds
to a certain order in the $1/m_Q$ expansion. We get
\begin{eqnarray}
\hat{R}_0^{(0)} (\omega) &=&  \bar{h}_v  \frac{1}
         {\omega + (i n \cdot D)}  \Gamma h_v
\\ \nonumber
\hat{R}_0^{(1)} (\omega) &=&  \bar{h}_v \frac{1}
         {\omega + (i n \cdot D)}  \Gamma h_v
 - \frac{1}{m_Q}
    \bar{h}_v \frac{1}{\omega + (in\cdot D)} (iD^\perp)^2 
              \frac{1}{\omega + (in\cdot D)} \Gamma h_v 
\\ \nonumber
\hat{R}_0^{(2)} (\omega) &=&  \bar{h}_v \frac{1}
         {\omega + (i n \cdot D)}  \Gamma h_v
 - \frac{1}{m_Q}
    \bar{h}_v \frac{1}{\omega + (in\cdot D)} (iD^\perp)^2 
              \frac{1}{\omega + (in\cdot D)} \Gamma h_v  \\ \nonumber
&& +  \frac{1}{4 m_Q^2}
    \bar{h}_v (i\fmslash{D}^\perp) \frac{1}
         {\omega + (i n \cdot D)} (i\fmslash{D}^\perp) \Gamma h_v \\ \nonumber
&& +  \frac{1}{4 m_Q^2}
    \bar{h}_v \left\{ (iD^\perp)^2 \, , \, \frac{1}
         {\omega + (i n \cdot D)} \right\} \Gamma h_v  \\ \nonumber 
&& + \frac{1}{m_Q^2}  \bar{h}_v  \frac{1}
         {\omega + (i n \cdot D)}
         (iD^\perp)^2  \frac{1}
         {\omega + (i n \cdot D)}
         (iD^\perp)^2  \frac{1}
         {\omega + (i n \cdot D)} \Gamma h_v 
\end{eqnarray}
where we have
\begin{equation}
\delta_R^\perp \hat{R}_0^{(k)} = {\cal O}(\Lambda_{QCD}^{k+3}/m_Q^{k+1})
\end{equation}
For the case $\Gamma = P_+$ we may reexpress
$\hat{R}_0^{(1)}$ in terms of the  $O_0 (\omega)$ and
$O_3(\omega_1, \omega_2)$ 
\begin{equation} \label{repara1}
\hat{R}_0^{(1)} (\omega) = 
\int \frac{d \sigma}{\omega - \sigma}  O_0 (\sigma)
- \frac{1}{2 m_Q} \int \frac{d \sigma_1}{\omega - \sigma_1}
                     \frac{d \sigma_2}{\omega - \sigma_2}
                     g_{\mu \nu} O_3^{\mu \nu} (\sigma_1, \sigma_2) 
\end{equation}
and replace $\omega \to \omega + i\epsilon$ in (\ref{repara1}) to 
identify
\begin{eqnarray}
&& R_0^{(1)} (\omega) = O_0 (\omega) - \frac{1}{\pi} {\rm Im} \left(
\frac{1}{2 m_Q} \int \frac{d \sigma_1}{\omega + i\epsilon - \sigma_1}
                     \frac{d \sigma_2}{\omega + i \epsilon - \sigma_2}
                     g_{\mu \nu} O_3^{\mu \nu} (\sigma_1, \sigma_2) \right)
\nonumber \\
&& =  O_0 (\omega) - 
\frac{1}{2 m_Q} \int d \sigma_1 \, d\sigma_2 \left(
\frac{\delta (\omega - \sigma_1) - \delta (\omega - \sigma_2)}
     {\sigma_1 - \sigma_2}\right)
                     g_{\mu \nu} O_3^{\mu \nu} (\sigma_1, \sigma_2)
\end{eqnarray}
to be the (up to order $\Lambda_{QCD}^4 /m_Q^2$) reparametrization-invariant
light-cone operator involving the leading order operator $O_0 (\omega)$.

Likewise, for $\Gamma = s^\alpha$ we get  
\begin{eqnarray}
Q_0^{\alpha (1)} (\omega) 
&=&  P_0^\alpha (\omega)  \\ \nonumber  
&& - \frac{1}{2 m_Q} \int d \sigma_1 \, d\sigma_2 \left(
\frac{\delta (\omega - \sigma_1) - \delta (\omega - \sigma_2)}
     {\sigma_1 - \sigma_2}\right)
                     g_{\mu \nu} P_3^{\mu \nu \alpha} (\sigma_1, \sigma_2)
\end{eqnarray}
for the spin-dependent reparametrization-invariant quantity up to order
$\Lambda_{QCD}^4/m_Q^2$.

The other operators of subleading order are not related to $O_0 (\omega)$
or $P_0^\alpha (\omega)$. 
In order to investigate the behaviour
$O_1^\mu (\omega)$, we split the covariant
derivative according to
\begin{equation} \label{covder}
iD^\mu = \frac{1}{2} (\bar{n}^\mu - n^\mu ) (i n \cdot D) + i D^\mu_\perp
\end{equation}
where we have made use of the equation of motion for the heavy quark, which
implies $ (i n \cdot D) = -  (i \bar{n} \cdot D)$ in $O_1^\mu (\omega)$ as
well as in $P_1^{\mu \alpha} (\omega)$.
In the same way as before we consider $\hat{O}_1^\mu (\omega)$ in which
the $\delta$ function is replaced by $1/(\omega+(i n \cdot D))$. According
to (\ref{covder}) we split $\hat{O}_1^\mu (\omega)$ into
$\hat{O}_{1,||}^\mu (\omega)$ and $\hat{O}_{1,\perp}^\mu (\omega)$. We get   
\begin{eqnarray}
\hat{O}_{1,||}^\mu (\omega) &=& (\bar{n}^\mu - n^\mu )
         \bar{h}_v (i n \cdot D) \frac{1}
         {\omega + (i n \cdot D)} \Gamma h_v \\ \nonumber
&=& (\bar{n}^\mu - n^\mu ) \left[\bar{h}_v \Gamma h_v - \omega  \bar{h}_v 
    \frac{1} {\omega + (i n \cdot D)} \Gamma h_v \right]
\end{eqnarray}
Taking the imaginary part (after $\omega \to \omega + i\epsilon$) we get
for $\Gamma = P_+$ 
\begin{equation}
O_{1,||}^\mu (\omega) = (n^\mu - \bar{n}^\mu )
                        \omega \bar{h}_v 
    \delta(\omega + (i n \cdot D)) h_v =  (n^\mu - \bar{n}^\mu)
                       \,  \omega \, O_0 (\omega) 
\end{equation}
which means that $O_{1,||}^\mu (\omega )$ is completely given in terms of
$O_0 (\omega)$. The same arguments apply for the spin-dependent operator
$P_1^{\mu \alpha} (\omega)$, where $P_{1 ||}^{\mu \alpha} (\omega)$ is
entirely given in terms of $P_0^\alpha (\omega)$

However, for the perpendicular pieces $O_{1,\perp}^\mu (\omega )$
and  $P_{1 \perp}^{\mu \alpha} (\omega)$ we get
for the reparametrization variation
\begin{eqnarray}
\delta_R^\perp O_{1,\perp}^\mu (\omega ) &=& - 2 m_Q \Delta^\mu_\perp
       O_0 (\omega)
   + {\cal O}(\Lambda_{QCD}^3/m_Q) \\
\delta_R^\perp P_{1,\perp}^{\mu \alpha} (\omega )
&=& - 2 m_Q \Delta^\mu_\perp P_0^\alpha (\omega)
   + {\cal O}(\Lambda_{QCD}^3/m_Q) 
\end{eqnarray}
which means that this variation contains a contribution of the same order
as the operator itself, which would need to be compensated by some other
subleading operator. However, there is no such operator, and so we conclude
that reparametrization invariance requires that only $O_{1,||}^\mu (\omega )$
and  $P_{1 ||}^{\mu \alpha} (\omega)$  contribute to a physical quantity. 

Using the same arguments we can discuss $O_2^\mu (\omega)$
and  $P_2^{\mu \alpha} (\omega)$. Obviously we have 
$O_{2 ||}^\mu (\omega) = 0 = P_{2 ||}^{\mu \alpha} (\omega)$.
However, unlike for $O_1^\mu (\omega)$ and  $P_1^{\mu \alpha} (\omega)$,
a reparametrization transformation yields 
\begin{eqnarray}
\delta_R^\perp O_{2,\perp}^\mu (\omega )  
   &=&  {\cal O}(\Lambda_{QCD}^3/m_Q) \\
\delta_R^\perp P_{2,\perp}^{\mu \alpha} (\omega )
   &=&  {\cal O}(\Lambda_{QCD}^3/m_Q) 
\end{eqnarray}
since these operators involve a commutator rather than an anticommutator,
and hence these operators will in general contribute.

Finally, nothing new can be obtained for
$O_4^{\mu \nu}$ as well as $P_4^{\mu \nu \alpha}$ from reparametrization
invariance; these operators are related through reparametrization to
higher order terms, which have not yet been classified. 

\section{Applications}
One immediate consequence of the above result concerns the matching
coefficients for light cone operators. Since physical observables
such as (differential) rates are reparametrization invariants, the
matching coefficients $C_0 (\omega)$ of $O_0 (\omega)$ and the one of
$O_3^{\mu \nu} (\omega_1, \omega_2)$ have to be related, such that
\begin{eqnarray} \label{raterep}
&& d \Gamma =  \int d \omega
              \left( C_0 (\omega) <R_0 (\omega)> + 
                     D_{0 \alpha} (\omega) <P_0^\alpha (\omega)> \right)
\\
&& = \int d \omega \left( C_0 (\omega) <O_0 (\omega)> +
    D_{0 \alpha} (\omega) <P_0^\alpha (\omega)> \right)
\nonumber \\ 
&& - \frac{1}{2 m_Q} \int d \omega C_0 (\omega)
     \int d \sigma_1 \, d\sigma_2 \left(
\frac{\delta (\omega - \sigma_1) - \delta (\omega - \sigma_2)}
     {\sigma_1 - \sigma_2}\right)
     g_{\mu \nu} < O_3^{\mu \nu} (\sigma_1, \sigma_2)> \nonumber \\
&& - \frac{1}{2 m_Q} \int d \omega D_{0 \alpha}  (\omega)
     \int d \sigma_1 \, d\sigma_2 \left(
\frac{\delta (\omega - \sigma_1) - \delta (\omega - \sigma_2)}
     {\sigma_1 - \sigma_2}\right)
     g_{\mu \nu} <P_3^{\mu \nu \alpha} (\sigma_1, \sigma_2)>  
\nonumber 
\end{eqnarray}
which can now be compared to (\ref{rate}), yielding the
reparametrization-invariance relation betwen the coefficients
\begin{eqnarray} \label{coeff}
C_{3,\mu \nu} (\sigma_1, \sigma_2) &=&  -\frac{1}{2} \int d \omega C_0 (\omega)
     \left(
\frac{\delta (\omega - \sigma_1) - \delta (\omega - \sigma_2)}
     {\sigma_1 - \sigma_2}\right)
     g_{\mu \nu} \\
C_{3,\mu \nu \alpha}^{(5)} (\sigma_1, \sigma_2) &=& -\frac{1}{2} \int d \omega
     C_{0, \alpha}^{(5)} (\omega)
     \left(
\frac{\delta (\omega - \sigma_1) - \delta (\omega - \sigma_2)}
     {\sigma_1 - \sigma_2}\right)
     g_{\mu \nu} 
\end{eqnarray}
Relation (\ref{coeff}) has been shown at tree level by explicit calculation
for the case of $B \to X_s \gamma$ in \cite{BLM1} and holds also for
the case $B \to X_u \ell \bar{\nu}_\ell$ \cite{BLM2},   
but here we claim that such a relation is a consequence of reparametrization
invariance and thus has to hold including radiative corrections.

In particular, it has to hold for the renormalization kernel of the
subleading operators $O_3^{\mu \nu} (\omega_1, \omega_2)$ and
$P_3^{\mu \nu \alpha} (\omega_1, \omega_2)$. While up to now only the
renormalization kernel of the leading order term has been investigated
\cite{BKM95,Aglietti}, a relation like (\ref{coeff}) has to relate the
kernel of $O_3^{\mu \nu} (\omega_1, \omega_2)$ with the one of 
$O_0 (\omega)$, and the kernel of $P_3^{\mu \nu \alpha} (\omega_1, \omega_2)$
will be related to the one of $P_0^\alpha (\omega)$. 

In this way, reparametrization invariance reduces the number of unknown
functions parametrizing e.g. the photon spectrum of $B \to X_s \gamma$
to subleading order. Following \cite{BLM1}, the non-vanishing matrix
elements leading to independent functions are
\begin{eqnarray}
\langle B(v) | O_0 (\omega) | B(v) \rangle &=& 2 m_B f(\omega) 
\\
\langle B(v) |O_3^{\mu\nu}(\omega_1,\omega_2) | B(v) \rangle &=&
2 m_B \, g_2 (\omega_1, \omega_2) g_\perp^{\mu\nu}
\nonumber \\
\langle B(v) | P_{2,\alpha}^\mu(\omega) | B(v) \rangle &=&
2 m_B\, h_1(\omega) \varepsilon_{\perp, \alpha}^\mu
\nonumber \\ 
\langle B(v) |P_{4,\alpha}^{\mu\nu}(\omega_1,\omega_2) | B(v) \rangle &=&
2 m_B\, h_2 (\omega_1, \omega_2)
\varepsilon_{\rho\sigma\alpha\beta} \, g_\perp^{\mu\rho}
g_\perp^{\nu\sigma}  v^\beta
\nonumber \\
\langle B(v) | O_T (\omega) | B(v) \rangle &=& 2m_B \,
t(\omega) \, ,
\nonumber
\end{eqnarray}
where we define
\begin{equation}
\varepsilon_\perp^{\mu \nu} =
\varepsilon^{\mu \nu \alpha \beta} v_\alpha n_\beta\,,
\end{equation}
and $\varepsilon^{0123} = 1$. Furthermore, $O_T (\omega)$ is the contribution
originating from the time-ordered product of the leading-order operator
$O_0 (\omega)$ with the $1/m_Q$ corrections to the Lagrangian; the precise
definition can be found in \cite{BLM1}. 

The contributions of $g_2$ and $h_2$ can be gathered into a fucntion of a
single variable 
\begin{eqnarray} 
G_2 (\sigma) &=& \int\! d\omega_1 \, d\omega_2 \, g_2(\omega_1, \omega_2)
\left[\frac{\delta(\sigma - \omega_1) - \delta(\sigma - \omega_2)}{\omega_1 -
\omega_2} \right]  \\
H_2 (\sigma) &=& \int\! d\omega_1 \, d\omega_2 \,
h_2(\omega_1, \omega_2) \left[\frac{\delta(\sigma - \omega_1) - \delta(\sigma -
\omega_2)}{\omega_1 - \omega_2} \right] \,.
\end{eqnarray}
which is at least for $g_2$ not surprising, since it is a consequence
of reparametrization invariance. In \cite{BLM2} the conclusion was reached
that the four universal functions
$F(\omega) = f(\omega) + t(\omega)/(2 m_Q)$, 
$G_2(\omega)$, $h_1(\omega)$ and $H_2 (\omega)$ are needed to parametrize
the subleading twist contributions to heavy-to-light decays. 

From reparametrization invariance we conclude that the functions $F(\omega)$
and $G_2 (\omega)$ have to appear always in the same combination, such that
\begin{equation}
{\cal F} (\omega) =  f(\omega) + \frac{1}{2 m_Q} t(\omega)
                               - \frac{1}{m_Q^2} G_2 (\omega) 
\end{equation}
is a single universal function. This has been confirmed at tree level
by explicit calculation, but this should hold to all orders in
$\alpha_s (m_b)$.  

\section{Conclusions}
We have discussed the consequences of reparametrization invariance for
the subleading contributions in the twist expansion for inclusive heavy
meson decays. As in the case for local operators, reparametrization
invariance relates different orders in the twist expansion. Looking at the
first subleading terms reparametrization relates the leading order shape
function to one of the subleading matrix elements, leading to identical
matching coefficients for the two contributions. As a practical concequence,
the spectra of inclusive heavy-to-light transitions are parametrized
in terms of three  unknown universal functions, once the first subleading 
terms are included.

\section*{Acknowledgements}
We are grateful to C. Bauer for comments on the manuscript. 
TM acknowledges the support of the Graduiertenkolleg ``Hochenergiephysik
und Teilchenastrophysik'' and of the Bundesministerium f\"ur Bildung und
Forschung. FC was supported by a stipend of the ``Marie Curie Program''
of the European Commission.

\end{document}